\documentstyle[12pt,aasms4,psfig]{article} 
\newcommand {\ks} {km~s$^{-1} \;$}

\slugcomment{Accepted by \apj}

\begin{document}
\title{NEARBY OPTICAL GALAXIES: \\
SELECTION OF THE SAMPLE AND IDENTIFICATION OF GROUPS}
\author{Giuliano GIURICIN$^{1, 2}$, Christian MARINONI$^{1, 4}$, 
Lorenzo CERIANI$^{1}$, Armando PISANI$^{3}$}
\affil{
$^1$ Dipartimento di Astronomia,
Universit\`{a} degli Studi di Trieste, \\
$^2$ SISSA, via Beirut 4, 34013 - Trieste, Italy, \\
$^3$ Osservatorio Astronomico, Trieste,\\
$^4$ Department of Physics, University of Waterloo, Waterloo, Canada \\
E-mail: giuricin@sissa.it; marinoni@stardust.sissa.it;\\
ceriani@mizar.sissa.it; pisani@ts.astro.it\\
}
\begin{abstract}

In this paper we describe the Nearby Optical Galaxy (NOG) sample, which is
a complete, distance-limited ($cz\leq$6000 km/s) and magnitude-limited
(B$\leq$14) sample of $\sim$7000 optical galaxies.  The sample covers 2/3
(8.27 sr) of the sky ($|b|>20^{\circ}$) and appears to have a good
completeness in redshift (98\%). We select the sample on the basis of
homogenized corrected total blue magnitudes in order to minimize
systematic effects in galaxy sampling. 

 We identify the groups in this sample by means of both the hierarchical and
the percolation {\it friends of friends} methods.  The resulting
catalogs of loose groups appear to be similar and are among the 
largest catalogs of groups presently available. Most of the NOG
galaxies ($\sim$60\%) are found to be members of galaxy pairs ($\sim$580
pairs for a total of $\sim$15\% of objects) or groups with at least three
members ($\sim$500 groups for a total of $\sim$45\% of objects). About
40\% of galaxies are left ungrouped (field galaxies). 

We illustrate the main features of the NOG galaxy distribution.  Compared
to previous optical and IRAS galaxy samples, the NOG provides a denser
sampling of the galaxy distribution in the nearby universe.  Given its
large sky coverage, the identification of groups, and its high-density
sampling, the NOG is suited for the analysis of the galaxy density field
of the nearby universe, especially on small scales. 
 
\end{abstract} 


\pagebreak

\section{INTRODUCTION} 

With the advent of large surveys of galaxy redshifts coupled to
well-selected galaxy catalogs, it has become possible to delineate the
three-dimensional (3D) distribution of galaxies and to attempt a
3D-definition of the galaxy density. 

This paper, which is the third in a series of papers (Marinoni et al. 1998, Paper I; 
Marinoni et al. 1999, Paper II) in which we investigate on the properties of the
large-scale galaxy distribution, presents the all-sky sample of optical galaxies
used in our study and the identification of galaxy groups in this sample. 

The first 3D galaxy catalogue which covered both Galactic hemispheres with
good completeness in redshift was the magnitude-limited (B$\lesssim$12 mag)
"Revised Shapley-Ames Catalogue of Bright Galaxies" (RSA, Sandage \&
Tammann 1981). It was used by Yahil, Sandage \& Tammann (1980) to calculate
the galaxy density field in the Local Supercluster (LS). The structures of
the LS region were well delineated by Tully \& Fisher (1987) on the basis
of the Nearby Galaxies Catalog (NBG, Tully 1988), which is a combination
of the RSA catalog and a diameter-limited sample of late-type and fainter
galaxies found in an all-sky HI survey. This catalog, which is limited to
a depth of 3000 km/s and is complete down to B$\sim$12 mag (although it
extends to fainter magnitudes), was also used to determine local galaxy
density parameters, which were exploited in statistical analyses of
environmental effects on some properties of the LS galaxies (Giuricin et
al. 1993, 1994, 1995; Monaco et al. 1994).

In an effort to go beyond the LS, Hudson (1993a, 1993b, 1994a, 1994b) 
constructed a wide galaxy sample from a merging of the diameter-limited 
northern UGC catalog (Nilson 1973) and the diameter-limited southern 
ESO catalog (Lauberts 1982; Lauberts \& Valentijn 1989). He applied 
statistical corrections for the fairly large incompleteness in redshift 
of his sample as a function of angular diameter and position on the sky 
and reconstructed the density field of optical galaxies to a depth of 
$cz=8000$ km/s.

The "Optical Redshift Survey" (ORS, Santiago et al. 1995), which provided
$\sim$1300 new redshifts for bright and nearby galaxies, marks a
considerable advance towards the construction of an all-sky sample of
nearby optical galaxies with good completeness in redshift. The ORS
contains $\sim$8300 galaxies with known redshift and consists of two
overlapping optically-selected samples (limited in apparent magnitude and
diameter, respectively) which cover almost all the sky with
$|b|>20^{\circ}$. Each sample is the concatenation of three subsamples
drawn from the UGC catalog in the north , the ESO catalog in the south
(for $\delta<-17.5^{\circ}$, and the Extension to the Southern Observatory
Catalogue (ESGC, Corwin \& Skiff 1999) in the strip between the UGC and
ESO regions ($-17.5\leq\delta\leq-2.5^{\circ}$). The authors selected
their own galaxy sample according to the raw (observed) magnitudes and
diameters and then attempted to quantify the effects of Galactic
extinction on the galaxy density field as well as the effects of random
errors and systematic trends in the magnitude and diameter scales internal
to different catalogs. They calculated the galaxy density field out to
$cz=8000$ \ks in redshift space on the basis of the UGC and ESO
magnitude--limited samples and of the ESGC diameter-limited sample (for a
total of $\sim$6400 galaxies), after having collapsed the galaxy members
of six rich nearby clusters to a single redshift (Santiago et al. 1996).
Baker et al. (1998) calculated the peculiar velocity field resulting from
the ORS sample (defined as above), adding the IRAS 1.2 Jy galaxy sample
(Fisher et al. 1995) in the unsurveyed ZOA ($|b|<20^{\circ}$) and at large
distances ($cz>8000$ km/s). 

In this paper, we follow a different approach to the construction of an all-sky
sample of optical galaxies with good properties of completeness, by attempting the use
of an uniform selection criterium (based on homogenized blue magnitudes corrected
for Galactic extinction, internal extinction and K-dimming) over the sky. The sample
we select (hereinafter Nearby Optical Galaxy (NOG) sample) is a complete,
magnitude--limited and distance--limited, all--sky sample of $\sim$7000 nearby and
bright optical galaxies, which we extract from the Lyon--Meudon Extragalactic
Database (LEDA) (e.g., Paturel et al. 1997).

This sample constitutes an extension in distance and in the number of redshifts
(with a consequent increase in redshift completeness) of the all-sky sample of
$\sim$6400 bright and nearby galaxies ($\sim$5400 galaxies above $|b|=20^{\circ}$),
recently used in the calculation of differents sets of galaxy distances corrected
for non--cosmological motions by means of peculiar velocity field models (Paper I)
and in the rediscussion of the local galaxy luminosity function (Paper II). 
                         
As previously emphasized (e.g., Hudson 1993a, Santiago et al. 1995), outside the
zone of avoidance (ZOA), optical galaxy samples are more suitable for mapping the
galaxy density field on small scales than IRAS--selected galaxy samples, which have
been frequently used as tracers of the galaxy density field on large scales (e.g.
Strauss et al. 1992, based on the IRAS 1.9 Jy sample, Fisher et al. 1995 and
Webster, Lahav \& Fisher 1997, both based on the IRAS 1.2 Jy sample, Branchini et
al. 1999 and Schmoldt et al. 1999, both based on the PSCz sample by Saunders et al. 
1999 a, b), because IRAS samples do not include the early-type galaxies (which have
little dust content and star formation), are relatively sparse nearby, and are based
on far-infrared fluxes, which are much less linked with galaxy mass than optical and
near-infrared fluxes. The latters are believed to be the best tracers of galaxy mass
and this motivates ongoing plans of constructing wide magnitude-limited samples of
near-infrared selected galaxies such as the 2MASS (e.g., Huchra et al. 1999) and
DENIS (e.g., Epchtein et al. 1999) projects. 

Moreover, as discussed by Santiago et al. (1996), standard extinction
corrections on diameters are thought to be less reliable than extinction
corrections on magnitudes. This makes it preferable to use
magnitude--limited optical samples rather than diameter--limited optical
ones for the reconstruction of the galaxy density field. 

Since we plan to use the NOG sample to trace the galaxy density field also on small
scales, in this paper we provide group assignments for the galaxies of the NOG
sample by means of both the hierarchical (H) (e.g., Tully 1987) and the percolation
(P) {\it friends of friends} methods (e.g., Huchra \& Geller 1982) of group
identification. The identification of groups, which allow us to study the continuity
of the properties of galaxy systems over a large range of scales (e.g.,  Girardi \&
Giuricin 1999), is also useful for improving the determination of the 3D structure
(e.g., the groups identified by Wegner, Haynes \& Giovanelli 1993 in the
Perseus--Pisces region). Furthermore, galaxy systems are favorite targets for
determining the peculiar velocity field with reduced uncertainties (e.g., Giovanelli
et al. 1997). In a forthcoming paper we shall use the locations of individual
galaxies and groups to reconstruct the galaxy density field (see Marinoni et al.
1999b for preliminary results). 

The outline of our paper is as follows. In \S 2 we present the selection
of the NOG sample. In \S 3 we illustrate the distribution of NOG galaxies
on the sky. In \S 4 we summarize the two identification procedures of
groups, i.e. the H and P algorithms. In \S 5 we present the resulting 
catalogs of loose groups. Conclusions are drawn in \S 6. 

Throughout, the Hubble constant is 75 km s$^{-1}$ Mpc$^{-1}$. 

\section{The selection of the sample}

Being aware that a sample must have a well-defined selection function in
order to be useful for any sort of quantitative work (e.g. the review by
Strauss 1999), we select a galaxy sample according to well-defined
selection criteria.  Relying, in general, on data (positions, redshifts,
total blue magnitudes) tabulated in LEDA, we select a sample of 7076
galaxies which satisfy the following selection criteria: 

\begin{itemize}
\item Galactic latitudes $|b|>20^{\circ}$; 
\item recession velocities (evaluated in the Local Group rest frame) 
$cz\leq$6000 km/s;
\item corrected total blue magnitudes B$\leq$14 mag. 
\end{itemize}

We transform tabulated heliocentric redshifts into the LG
rest frame according to Yahil, Tammann \& Sandage (1997). In the following
we always refer to redshifts evaluated in the LG frame. 

Limiting the sample to a given depth ($cz\leq$6000 km/s in our case) has
the main advantage of reducing the incompleteness in redshift for a given
limiting magnitude, because a fraction of the galaxies with unknown
redshift is presumably located beyond the limiting depth. With this choice
our sample is also less affected by shot noise which increases with
increasing distance. Moreover, the choice of limiting the volume of the
sample minimizes distance effects in the identification of galaxy groups.
Last, the knowledge of the peculiar velocity field, which will be used to
place the NOG objects into the real-distance space, becomes very poor
beyond this depth. 
 
In the LEDA compilation, which collects and homogenizes several data for
all the galaxies of the main optical catalogues --- such as the catalogs
UGC, ESO, ESGC, CGCG (Zwicky et al. 1961--1968) and MCG
(Vorontsov--Velyaminov, Archipova \& Krasnogorskaja 1962--1974)---, the
original raw data (blue apparent magnitudes and angular sizes) have been
transformed to the standard systems of the RC3 catalog (de Vaucouleurs et
al. 1991) and have been corrected for Galactic extinction, internal
extinction, and K-dimming, as described in Paturel et al. 1997.
Corrections for internal extinction, which are conspicuous in very
inclined spiral galaxies, are in general neglected in magnitude-limited
optical galaxy samples used in studies of the spatial galaxy
distributions. The adopted corrections for internal extinction do not 
take into account a possible dependence on the galaxy luminosity (e.g., 
Giovanelli et al. 1995).

The adopted limits for the unsampled ZOA ($|b|<20^{\circ}$) are imposed by the
requirement of intrinsic completeness of the sample. An additional problem
which affects the construction of a well-controlled optical galaxy sample in
the ZOA is the presumably low quality of available Galactic reddening maps in
this region. As a matter of fact, precisely in the ZOA there are pronounced
differences between the classical maps of Burstein \& Heiles (1978, 1982)
(substantially adopted in LEDA), which are largely HI maps with the zero-point
adjusted and with smooth variations in dust-to-gas ratio estimated from galaxy
counts, and the new maps derived by Schlegel, Finkbeiner \& Davis (1998) from
the COBE/DIRBE and IRAS/ISSA observations, which give a direct measure of the
column density of the Galactic dust. Tests of the accuracy of reddening maps
emphasize their unreliability in regions characterized by a strong and very
patchy Galactic extinction (e.g. Arce \& Goodman 1999) such as the low
$|b|$-regions and reveal large-scale errors across the sky in the ZOA,
specifically an appreciable overestimate of Galactic extinction in the Vela
region ($230^{\circ}<l<310^{\circ}$, $|b|<20^{\circ}$) (Burstein et al. 1987; 
Hudson 1999). 

In the LEDA there are 6880 galaxies which satisfy the adopted selection
criteria (B$\leq$14 mag, $cz\leq$6000 km/s, $|b|>20^{\circ}$).  We add to
this initial sample 196 galaxies (with B$\leq$14 and $|b|>20^{\circ}$)
which have new measures of redshifts that we find from matching the LEDA
with the NASA Extragalactic Database (NED), the Updated Zwicky
Catalog (UZC) (Falco et al. 1999), the ORS (Santiago et al. 1995) and the 
PSCz (kindly provided to us by B. Santiago and W. Saunders, respectively).  

Relying on information given in LEDA and NED for the binary and multiple
systems of galaxies, we include in our sample only the individual
components in these systems which satisfy our selection criteria. 
 
The final distance-limited ($cz\leq$6000 km/s) and magnitude-limited (B$\leq$
14 mag) NOG sample comprises 7076 galaxies (with $|b|>20^{\circ}$).
 
The logarithmic integral counts of all LEDA galaxies versus their blue
total magnitude show a linear relation down to $\sim$15.5 mag (Paturel et
al. 1997), whilst the logarithmic differential counts of all LEDA galaxies
with $|b|>20^{\circ}$ reveal that a linear relation is satisfied only down
to magnitudes somewhat fainter than B=14 mag, which can be regarded as the
limit of intrinsic completeness of the data base.
   
Thus, although the different galaxy catalogues, from which data are
collected and homogenized in the LEDA, have different limits of
completeness in apparent magnitude or angular diameter, the NOG sample
turns out to be nearly intrinsically complete down to its limiting
magnitude $B=14$ mag.  

The redshift completeness of all-sky samples of bright optical galaxies is
not yet extremely high and decreases with fainter limiting magnitudes 
(e.g. Giudice 1999). For the sample limited to $|b|>20^{\circ}$ and
B$\leq$14 mag there are 550 objects without redshift measures. Some of
these objects are galaxies with bright stars superposed for which is
difficult to obtain a spectrum. Most of these objects are galaxies with
faint (uncorrected) apparent magnitudes. Most of the objects 
without redshift are located in the southern sky 
(precisely at $\delta<-10^{\circ}$).
 
Thus, the degree of redshift completeness of this sample, with no
limits in redshift, is 92\%. This is indeed a lower limit to the redshift
completeness of the NOG, since the NOG is limited to 6000 km/s. 

We have estimated the NOG redshift completeness C by dividing the number
$N_z$ of galaxies with known redshift ($N_z$=7076) by the total number
$N_T=N_z+N_p$ of galaxies which are presumed to have $cz\leq$6000 km/s. We
have calculated the number $N_p$ of objects with unknown redshifts which
are predicted to have $cz\leq$6000 km/s as $N_p=\sum_{i=1}^{n} P_i(B)$,
where $P_i(B)$ is the probability for a galaxy with magnitude B and
unknown redshift to have $cz\leq$6000 km/s. We have estimated this
probability under the assumption of a homogeneous universe for the
Schechter-like galaxy luminosity function which fits the differential
galaxy counts. In this way we obtain a redshift completeness of 98\%,
which is a fixed average percentage over the sampled volume. Details on 
these calculations and on the selection function of the NOG sample 
will be presented in a subsequent paper (see Marinoni et al. 1999b  
for preliminary results).

Adopting a sample selection based on corrected and homogenized magnitudes,
we attempt to minimize systematic selection effects as a function of
direction in the sky, which may arise from inconsistencies among the
different magnitude systems used in the original catalogs, and we take
into account the variable amounts of Galactic extinction across the sky
and of internal extinction in galaxies of different morphological types
and inclination angles. Clearly, systematic errors (though not zero-point 
errors in Galactic and internal extinctions) across the sky would affect the 
uniformity of galaxy sampling. 
 
Notwithstanding the different selection criteria adopted, the NOG sample
has many galaxies in common with the ORS sample, which comprises
$\sim$6280 galaxies having $cz\leq$6000 km/s (and $|b|>20^{\circ}$), of
which $\sim$4360 and $\sim$4280 objects belong to the magnitude-limited
and diameter-limited ORS subsamples, respectively. A large fraction of
these galaxies, 87\% (95\% and 86\% of those belonging to the
magnitude-limited and diameter-limited ORS subsamples restricted to
$cz\leq$6000 km/s), are common to the NOG. There are 78\% of NOG galaxies
common to the ORS; to be more precise, 59\% and 52\% of NOG galaxies are
common to the magnitude-limited and diameter-limited ORS subsamples,
respectively. 

\section{The distribution of galaxies on the sky} 

Fig. 1 shows the distribution of  NOG galaxies on the celestial sphere
using equal--area Aitoff projections in equatorial, Galactic, and 
supergalactic coordinates. The region devoid of galaxies 
corresponds to the unsampled ZOA ($|b|<20^{\circ}$).

Although Galactic extinction is greater than the norm in the center
($l\sim0^{\circ}$) and anticenter ($l\sim180^{\circ}$) regions, there may be a real
deficiency of galaxies in these regions at low $|b|$-values. In particular, this is
suggested by redshift surveys which select galaxy candidates from the IRAS Point
Source Catalog (1988), whose completeness is, however, quite questionable in these
two regions. Specifically, a concatenation of large voids stretching from the Local
Group all the way to the NOG distance limit and beyond (see, e.g., Lu \& Freudling
1995) is thought to be responsible for the deficiency of galaxies in the
Orion--Taurus anticenter region ($l=150^{\circ}-190^{\circ}$, $b\sim-30^{\circ}$). 
As regards the center region, redshift surveys have pointed out the presence of a
nearby void, around $l=0^{\circ}$ and $b=10^{\circ}$, the Ophiuchus void (Wakamatsu
et al. 1994, Nakanishi et al. 1997). This void appears to be contained in the large
Local Void of Tully \& Fisher (1987), which covers a large part of the sky between
$l\sim0^{\circ}$ and $l\sim80^{\circ}$.  The Local Void, which is centered at
$cz\sim$2500 km/s and has a diameter of $\sim$2500 km/s (Nakanishi et al. 1997), is
probably interconnected with the more distant, large Microscopium void (centered at
$b\sim0^{\circ}$, $l\sim10^{\circ}$, $cz\sim$4500 km/s). 

In order to distinguish structures more clearly, in Fig.2 we show the
Aitoff projections of the NOG galaxies on the celestial sphere in Galactic
coordinates, for three redshift slices. 

Prominent structures stand out in these plots. Many galaxies tend
to be concentrated in the supergalactic plane which stretches in the plots
from $l\sim135^{\circ}$ to $l\sim315^{\circ}$. The densest part of the
Local Supercluster is the overdensity at $l=300^{\circ}-315^{\circ}$,
$b=30^{\circ}-70^{\circ}$ (Virgo Southern Extension) with the Virgo
cluster at its northern tip ($l=284^{\circ}$, $b=75^{\circ}$).

In the low-redshift slice ($cz<$2000 km/s, 2012 galaxies) we further note some
nearby clusters, such as Ursa Major ($l=145^{\circ}$, $b=66^{\circ}$), Fornax
($l=237^{\circ}$, $b=-54^{\circ}$), and the cluster surrounding NGC 1395 in the
Eridanus cloud ($l=214^{\circ}$, $b=-52^{\circ}$). The last two clusters are
the dominant overdensities of the Dorado--Fornax--Eridanus complex, also named
Fornax wall (the southern supercluster of Mitra 1989), which ranges from
$l=190^{\circ}$, $b=-60^{\circ}$ to $l=270^{\circ}$, $b=-40^{\circ}$.  The
Local Void is apparent as the paucity of galaxies between $l\sim0^{\circ}$ and
$l\sim80^{\circ}$.  Other voids are discernible, e.g. the Gemini void around
$l=190^{\circ}$, $b=20^{\circ}$. The latter void is a part of a very large
nearby void (named V$\alpha$ by Webster et al. 1997) which stretches below the
Galactic plane down to the above-mentioned Orion-Taurus void
($l=150^{\circ}-190^{\circ}$, $b\sim-30^{\circ}$). 

The intermediate-redshift slice ($2000\leq cz<$ 4000 km/s, 2377 galaxies)
intersects the Great Attractor region, which includes the Hydra--Centaurus
complex, which stands out around $b=20^{\circ}$, $l=260^{\circ}$ (Hydra)
and $l=310^{\circ}$ (Centaurus), together with the contiguous
Telescopium--Pavo--Indus (T--P--I) supercluster (also named Centaurus Wall), 
whose foreground part is apparent from $b=-20^{\circ}$, $l=330^{\circ}$ to
$b=-60^{\circ}$, $l=30^{\circ}$, and the Hydra Wall, which starts from the
Hydra cluster and stretches in the southern Galactic hemisphere from
$b=-20^{\circ}$, $l=230^{\circ}$ to $b=-30^{\circ}$, $l=190^{\circ}$. 
Noticeable clumps in the northern hemisphere are the Canes
Venatici--Camelopardalis clouds at $l=95^{\circ}$,
$50^{\circ}<b<70^{\circ}$ and the Ursa Major cloud at $l=130^{\circ}$,
$30^{\circ}<b<60^{\circ}$.  There is a prominent void, the Leo void, at
$l\sim200^{\circ}$, $b\sim60^{\circ}$. The large Eridanus void around
$l=270^{\circ}$, $b=-60^{\circ}$, which roughly corresponds to the void
named V1 da Costa et al. (1988) and V$\beta$ by Webster et al. (1997),
stretches considerably towards the Galactic plane. 

In the next redshift slice ($cz\geq$4000 \ks, 2653 galaxies) the dominant
overdensities are the Perseus--Pisces supercluster
($l=110^{\circ}-150^{\circ}$, $-35^{\circ}<b<-20^{\circ}$) and the main
part of the Telescopium--Pavo--Indus supercluster in the southern Galactic
hemisphere. The Cetus Wall runs southwards from Perseus--Pisces along
$b\sim-60^{\circ}$. The galaxy concentration around $l=190^{\circ}$,
$b=-25^{\circ}$ is the NGC 1600 region (Saunders et al. 1991). The galaxy
overdensity around $l=120^{\circ}$, $b=-70^{\circ}$, which does not
correspond to a specific galaxy cluster, was named C$\gamma$ by Webster
et al. (1997). The void at $l=300^{\circ}$, $b=-45^{\circ}$ was named V3
by da Costa et al. (1988). In the northern sky we recognize the
high-redshift component of the Hydra-Centaurus complex with the
surrounding Hydra void (at $l\sim290^{\circ}$, $b\sim30^{\circ}$), the
Cancer cluster ($l=195^{\circ}$, $b=25^{\circ}$), the Gemini filament (at
$180^{\circ}<l<210^{\circ}$, $15^{\circ}<b<30^{\circ}$; see Focardi,
Marano \& Vettolani 1986), the Cygnus-Lyra filament (see Takata, Yamada \&
Sait\=o 1996) which crosses the Galactic plane from $l\sim90^{\circ}$,
$b\sim15^{\circ}$ to $l\sim50^{\circ}$, $b\sim10^{\circ}$, and the
Camelopardalis supercluster ($l=135^{\circ}$, $b=25^{\circ}$), which,
according to Webster et al.  1997, is probably connected with the
Perseus-Pisces supercluster.  

The large void which covers most of the northern sky between
$l=145^{\circ}$ and $l=195^{\circ}$ lies between the Virgo cluster and the
"Great Wall" and was noted in the CfA1 redshift survey of Davis et al.
(1982).  

In Figs. 3 and 4 we show the distribution of NOG galaxies on the celestial
sphere using equal-area polar hemispheric projections in equatorial
coordinates, for different redshift slices. These plots better illustrate
many other minor structures and voids in the galaxy distribution. The
structures illustrated in our plots are qualitatively similar to those
described in the analogous plots presented in Fairall' s (1998) books for
a generic (statistically uncontrolled) wider sample of galaxies with known
redshift and no limit in magnitude (or diameter). This book gives a
comprehensive description of the cosmography of the nearby universe (see
also Tully \& Fisher 1987 and Pellegrini et al. 1990, for previous
detailed descriptions of the structures of the Local Supercluster and
southern hemisphere, respectively).

The distribution of NOG galaxies appears qualitatively similar to that of
the ORS galaxies ({\it cfr} the analogous Aitoff projections presented by
Santiago et al. 1995 and Baker et al. 1998). Both optical galaxy samples
trace essentially the same structures, with NOG providing a somewhat
denser sampling (11\% more galaxies) of the galaxy density field in the
nearby universe (within 6000 \ks). Moreover, comprising 3204 galaxies with
$cz\leq$3000 km/s, the NOG gives a much denser sampling of the LS region
than the NBG sample. 

A comparison with the distribution of the IRAS 1.2--Jy galaxies ({\it cfr}
the plots given by Fisher et al. 1995 and Baker et al. 1998) shows that NOG
samples the galaxy density field much better than the IRAS samples and
delineates similar major overdensity regions but with a greater density
contrast. This is related to the known fact that IRAS surveys under-count
the dust-free early-type galaxies which congregate in high-density regions
and give a galaxy density field characterized by a bias smaller by a
factor of $\sim1.5$ than that of the optical galaxy density field (e.g.,
Strauss et al. 1992; Hudson 1993; Hermit et al. 1996). The newly completed
PSCz survey (Saunders et al. 1999 a, b), which includes IRAS galaxies to a flux
limit of 0.6 Jy, leads to a density field which compares fairly well with
that derived from the IRAS 1.2 Jy sample (e.g., Branchini et al. 1999; 
Schmoldt et al. 1999). Although the NOG covers 79\% of the solid angle 
covered by the PSCz, our sample contains 35\% more galaxies.

\section{The Identification of Galaxy Groups}.

We identify galaxy groups by means of the most widely used objective
group-finding algorithms, the hierarchical and the percolation {it\
friends of friends} algorithms, which allow us a comparison with wide
group catalogs published in the literature, although other objecting
techniques of clustering analysis are available (e.g., Pisani 1996, 
Escalera \& MacGillivray 1995). 

\subsection{The hierarchical algorithm}

In the hierarchical (H) clustering method, first introduced by Materne 
(1978), one defines an affinity parameter between the galaxies (e.g. their
separations) which controls the grouping operation. Then one starts with
all galaxies of the sample as separate units and links the galaxies
successively in order of affinity until there is only one unit that
encompasses the ensemble.  A hierarchical sequence of units organized by
decreasing affinity is the result of this method. The merging of a galaxy
into a given unit involves the consideration of the whole unit and not
only of the last object merged into the unit. Another merit of this method
is the easy visualization of the whole merging procedure under the form of
a hierarchical arborescence, the dendrogram. 

Customarily, it is believed that the H method has the practical drawback
of requiring very long calculation time (e.g., in comparison with the
percolation method). Paying attention to this problem, we have managed to
considerably speed up the hierarchical code by using numerical tricks. In
this way, we have made this code nearly as fast as the percolation
algorithm. The code is written in the C programming language, which allows
us to use techniques of sparse matrix (i.e with most elements equal to
zero) in a natural way, through a data structure based on pointers.
Specifically, for each pair of NOG galaxies, the affinity parameter, which
is taken to be the galaxy luminosity density as explained below, is not
stored in memory and is not exactly calculated, but replaced with zero, if
its value is smaller than a preselected limit. The maximum value of this 
parameter is searched only for the few pairs for which the parameter 
values are greater than this limit. Then the limit is gradually lowered  
in the following steps until the dendrogram is completed.

There are several possible choices for the grouping parameter. For
instance Tully (1980, 1987) and Vennik (1984) employed a grouping
parameter (galaxy luminosity divided by separation squared) which measures
the gravitational force between galaxies $i$ and $j$, but cut the
hierarchy according to the luminosity density and number density of the
entity, respectively. 

Following basically the procedure adopted by Gourgoulhon, Chamaraux \&
Fouqu\'e (1992), we use the same parameter for the two operations, namely
the luminosity density $3 (L_i + L_j)/(4 \pi r_{ij}^{3})$, where $L_i$ and
$L_j$ are the corrected luminosities (as defined below) of the galaxies
$i$ and $j$, and $r_{ij}$ their mutual separation.  We take into account
the loss of faint galaxies with increasing distances within our
magnitude-limited galaxy sample by multiplying the luminosity of each
galaxy located at a distance $r$ by the factor
 
\begin{equation}
\beta(r)= \frac{\int_{0}^{\infty} L \Phi(L) dL}{\int_{L_{min}(r)}^{\infty} 
L \Phi(L) dL}
\end{equation}

\noindent where $\Phi(L)$ is the galaxy luminosity function of our 
sample, $L_{min}$ is the minimum luminosity necessary for a 
galaxy at a distance $r$ (in Mpc) to make it into the sample; $L_{min}$ 
corresponds to the absolute magnitude $M_B= -5 \log r - 25 + B_{lim}$, 
where $B_{lim}$=14 mag is the limiting apparent magnitude of our sample.

We use the Schechter (1976) form of the luminosity function with $M^{*}=-20.68$,
$\alpha=-1.19$, $\Phi^{*}=0.0052 Mpc^{-3}$. This is the luminosity function,
unconvolved with the magnitude error distribution (i.e., not Malmquist-corrected,
according to the precepts of Ramella, Pisani \& Geller 1997), that we derive by
means of Turner's (1979) method (see also de Lapparent, Geller \& Huchra 1989 and 
Paper II).  For this calculation, using redshifts as distance
indicators, we take the the NOG galaxies having $cz>500$ km/s, and $M_B$-values in
the range $-22.5\leq M_B\leq M_l$, where $M_l= -15.12$ is the faintest absolute
magnitude at which galaxies with magnitude limit $B_{lim}$=14 mag are visible at the
fiducial distance $r=500/(75\cdot h_{75})\sim 6.7\cdot h_{75}^{-1}$ Mpc.  Convolving
the Schechter form of the luminosity function with a Gaussian magnitude error
distribution having zero mean and dispersion of 0.2 mag, we obtain the
Malmquist-corrected luminosity function characterized by $M^{*}=-20.59\pm0.07$,
$\alpha=-1.16\pm0.05$, $\Phi^{*}=0.0065\pm 0.0009\; Mpc^{-3}$. The 
luminosity function is similar to that derived in Paper II from
a similar, albeit smaller and less complete in redshift, sample of nearby and bright
optical galaxies (see Paper II for a detailed discussion and comparison 
with the galaxy luminosity functions given in the literature).   

For $B_{lim}$=14 mag, $\beta$ is 1.19, 1.74, 3.07 at 2000, 4000, 6000 km/s 
respectively.

We adopt $8\cdot 10^{9} L_{\odot} Mpc^{-3}$ (corresponding to a luminosity density
contrast of 45) as the limiting luminosity density parameter used to cut the
hierarchy and define groups. The same value was adopted by Gourgoulhon et al.
(1992). Tully (1987), using only the luminosity of the brighter component in the
evaluation of the entity density, chose the slightly smaller value of $2.5\cdot
10^{9} L_{\odot} Mpc^{-3}$. We have checked that the value adopted by us better
distinguish some known nearby structures, such as the substructures identified in
the Virgo cluster region by specific surveys (see end of this subsection), than
Tully's (1987) value does. 

Following Tully (1987) and Gourgoulhon et al.  (1992), we distinguish two
cases in the derivation of the separation $r_{ij}$ between galaxies $i$
and $j$ from their angular distance.  In the case of small differences in
the velocities, we assume that no information is available about the
line--of--sight separations in differential velocities and take
separations from plane--of--sky information, with the average projection
factor $4/\pi$ applied to correct statistically for depth in the third
dimension (see eq. 4 in Gourgoulhon et al. 1992). 

In the case of large differences in the velocities , we assume 
that differential velocities are simply related to the expansion of the 
universe and directly infer a line--of--sight separation (see eq. 7 
in Gourgoulhon et al. 1992). 

For intermediate cases, we use the transition formula proposed by 
Gourgoulhon et al. (1992) (see their eqs. 5 and 6), which transforms 
between the two above-mentioned limiting cases in a smooth way.

The procedure is regulated by the choice of a free parameter, the transition
velocity $V_l$. The choice of $V_l$ is a compromise between too low values which
would lead to rejection of group members with large peculiar velocities (with a
consequent underestimate of the group velocity dispersion) and too high values
which would allow the inclusion into groups of galaxies which are accidental
superpositions in the line of sight (with a consequent overestimate of the group
velocity dispersion).  Following Gourgoulhon et al. (1992), we adopt the fairly
low value of $V_l$=170 km/s, which reliably identifies groups of low velocity
dispersion.  For his less deep sample, Tully's (1987) choice, $V_l$=300 km/s, was
greater than our value; moreover, his value is roughly equivalent (in terms of
corresponding galaxy separations) to the value we adopt, in view of the different
transition formula employed by this author. 

With low values of $V_l$ the clusters of galaxies are split into various
subunits because of their large velocity dispersion.  These subunits are
located at about the same positions, but have different average
velocities. This inconvenience of the method is related to the use of an
universal $V_l$-value for the whole sample. 

As done by Gourgoulhon et al. (1992), after running the algorithm, we
identify by hand 17 high-velocity, relatively rich systems, by collecting
the various subunits into one aggregate (for a total of 440 galaxies), with 
the aid of the results obtained with the P algorithm (in the variant P1) 
discussed in \S 4.2. Tully (1987) removed the high-velocity systems before 
running the algorithm, which implies that system members are to be 
chosen a priori, whilst Garcia (1993) neglected this problem in many 
cases.

There are two regions of the sky where the initial results obtained from
running the H algorithm were unsatisfactory, i.e. the region
comprising the nearest systems to the Local Group and the complex region
of the Virgo cluster. In the former case the algorithm groups together
many nearby galaxies, because the redshift is no longer a reasonable
indicator of distance; in this case, reliable results could be obtained
from the algorithm by replacing the redshifts with redshift-independent
distances. Therefore, to identify very nearby systems, we have
first selected the members of four well-known nearby groups directly on
the basis of the specific studies by van Driel et al. (1998) for the M81
group, by C\^{o}t\'e et al. (1997) for the Sculptor and Centaurus A
groups, on the review by Mateo (1998) for the Local Group. Then, after
having excluded the members of these groups, we have rerun the algorithm
for the other galaxies.  

Since a long time specific surveys of the Virgo region have identified
substructures in the Virgo cluster first by means of an inspection of the
morphological classification, brightness, redshift of galaxies (e.g.,
Binggeli, Sandage \& Tammann 1985) and then through accurate distance
indicators (mainly the Tully -- Fisher relation for spirals). The current
knowledge of the main clumps of the Virgo cluster, which appears to be a
structure considerably elongated along the line of sight, can be
summarized as follows (see, e.g., the recent studies by Yasuda, Fukugita
\& Okamura 1997, Federspiel, Tammann \& Sandage 1998, Gavazzi et al.
1999):  the subcluster A centered on the galaxy M87 is the dominant
substructure (at a velocity $cz\sim$1350 km/s and at a distance of
$\sim$14-18 Mpc); the clump B, offset to the south around M49, lying at
similar cz but at larger distance ($\sim$20-24 Mpc), is thought to fall to
Virgo A; the clouds M, W (both at $cz\sim$ 2500 km/s) are background
structures at twice the distance of Virgo A and may also be falling to
Virgo A; the cloud W' is located at $cz\sim$1500 km/s and $\sim$25 Mpc;
the northern part of the Virgo Southern Extension (SE) lies at a redshift
and distance similar to that of the main body.  In this paper we have made
membership assignments adopting borderlines between the different
substructures in accordance with Binggeli, Tamman \& Sandage (1987) and
Binggeli, Popescu \& Tammann (1993). 

\subsection{The {\it friends of friends} algorithm}

We identify groups in redshift-space with the percolation (P) {\it friends
of friends} algorithm (Huchra \& Geller 1982). So far, this algorithm,
being easier to implement than the H algorithm, has been the most widely
used method of group identification in the literature.  Unlike the H
algorithm, this algorithm does not rely on any {\it a priori} assumption
about the geometrical shape of groups, although it may suffer from some
drawbacks which are mentioned at the end of \S 4.2. 

For each galaxy in the NOG sample, this algorithm identifies all other
galaxies with a projected separation $D_{12}\leq D_L(cz_1, cz_2)$ and a
line-of-sight velocity difference $cz_{12}\leq cz_L(cz_1,cz_2)$ where
$cz_1$, $cz_2$ are the velocities of the two galaxies in the pair. All
pairs linked by a common galaxy form a group. We estimate the limiting
number density contrast as

\begin{equation}
\frac{\delta \rho}{\rho} = \frac{3}{4 \pi D_{0}^{3}} 
\bigg[\int_{\infty}^{M_l} \Phi(M) dM\bigg]^{-1} - 1
\end{equation}

\noindent where $\Phi(M)$ is the luminosity function of the sample (see \S
5.1) and $M_{l}= -15.12$ mag is the faintest absolute magnitude at which
galaxies with magnitude limit $B$=14 mag are visible at the fiducial
distance $r=500/(75\cdot h_{75})\sim 6.7\cdot h_{75}^{-1}$ Mpc. The
estimate assumes that the galaxy separation along the line of sight is
comparable with $D_L$ (e.g., spherical symmetry). 

In order to take into account the decrease of the magnitude range of the
luminosity function sampled at increasing distance, the distance link
parameter $D_L$ and the velocity link parameter $cz_L$ are in general
suitably increased with increasing distance. Huchra \& Geller 
(1982) initially and later other authors (e.g., Geller \& Huchra 1983;  Maia,
da Costa \& Latham 1989; Ramella, Geller \& Huchra 1989; Ramella, Pisani
\& Geller 1997) scaled the distance and velocity link parameters in the
same way, as $D_L=D_0\cdot R$ and $cz_L=cz_0\cdot R$, where

\begin{equation} 
R=\biggl[\int_{\infty}^{M_l} \Phi(M) dM / \int_{\infty}^{M_{12}} \Phi(M) 
dM\biggr]^{1/3}
\end{equation}

\noindent
and $M_{12}$ is the faintest absolute magnitude  at which a galaxy  
with apparent magnitude equal to the magnitude limit ($B=14$ mag in our case) 
is visible at the mean distance of the pair. Scaling both $D_L$ and $cz_L$ with 
distance, one keeps the number density enhancement, $\delta \rho/\rho$, constant. 

The properties of selected groups are known to be sensitive to the adopted
distance and velocity links. As a matter of fact, the typical size of a
group is mostly linearly related to the adopted value of $D_0$, whereas
the typical velocity dispersion of a group mostly depends on the adopted
value of $cz_0$ (e.g., Trasarti-Battistoni 1998).  The adopted value of
$cz_L$ must be small enough to avoid the inclusion of too many interlopers
in groups, without biasing the velocity dispersion of groups towards too
low values.  The chosen value of $\delta \rho/\rho$ must be large enough
to avoid that unbound fluctuations in the distribution of galaxies within
large scale structures be mistaken for real systems, without splitting
rich systems into many multiple systems. 

Geometrical Monte-Carlo simulations (Ramella et al. 1989, 1997) and especially
cosmological N-body simulations which have used full 3D information (e.g.,
Nolthenius \& White 1987; Moore, Frenk \& White 1993; Nolthenius, Klypin \& Primack
1994; Frederic 1995 a, b; Nolthenius, Klypin \& Primack 1997; Diaferio et al. 1999)
can help us in searching for the optimal sets of linking parameters and scaling
relations with distance which maximize the efficiency of the P algorithm in picking
up "real" groups. As a matter of fact, almost all relevant simulations were designed
to describe the properties of redshift surveys whose magnitude limits are comparable
to that of NOG (e.g., CfA1) or moderately fainter than that of NOG (e.g., CfA2),
which, however, is limited to a smaller distance .  Moreover, moderate differences
in the luminosity functions and magnitude limits of galaxy samples (e.g. CfA1 versus
CfA2) lead to minor differences (on the order of 10-15\%) in the optimal choices of
percolation linking parameters (as discussed by Trasarti-Battistoni 1998). 

Investigations on the variation of the properties of groups (identified
in several redshift surveys) with $cz_0$ and $D_0$ (or $\delta
\rho/\rho$) showed that there is a range of values of the two 
parameters where the median properties of the groups are fairly stable
(i.e., $\delta \rho/\rho=$60--160, $cz_L=$200--600 km/s at the velocity
of 1000 km/s), with an "optimal choice" believed to be centered around
$\delta\rho/\rho$=80 and $cz_L$=350 km/s (at the velocity of 1000 km/s)
(e.g. Ramella et al. 1989, 1987; Frederic 1995 a, b). These simulations
also show that an appreciable fraction of the poorer groups, those with
$n<5$ members, is false (i.e. unbound density fluctuations), whereas the
richer groups almost always correspond to real systems.
 
More specifically, testing the accuracy of group-finding algorithms
through N-body cosmological structure simulations, Frederic (1995 a, b)
pointed out that the optimal parameters which maximize the accuracy of
group identification are indeed dependent on the purposes for which
groups are being selected.  With the above-mentioned scaling of the
linking parameters, restrictive velocity linking lengths (i.e.,
$cz_L\sim$200 km/s at 1000 km/s) tend to cause members of the few high
velocity dispersion systems to be missed (biasing low their velocity
dispersion and mass), but result in a much fewer interlopers. Therefore
generous velocity links (i.e., $cz_L\sim$500 km/s at 1000 km/s) may be
preferred in studies which aim to well identify high-velocity dispersion
systems; on the other hand, restrictive velocity links, which is what we
will choose in this paper, are to be preferred in our case, because the
NOG is limited to a relatively small depth and (unlike the CfA1 and CfA2
samples) it does not contain very rich (e.g. Coma-like) galaxy clusters
and especially because we shall use the NOG groups mainly to collapse
their members to a single redshift, removing peculiar motion effects on
group scales. Consistently with these considerations, Nolthenius (1993),
who revised the identification of CfA1 groups with the introduction of
galaxy distances calculated from a Virgo-Great Attractor flow field
model, reduced significantly the interloper contamination by choosing a
restrictive velocity link ($cz_L$=350 km/s at 5000 km/s, i.e. a value of
$cz_L$ only $\sim$ 1/4 as large as that chosen in the original catalog
of CfA1 groups by Geller \& Huchra 1983). 

We have run the P algorithm (with the above-mentioned scaling of $D_L$ and
$cz_L$) for some pairs of values of the two linking parameters in the
above-mentioned ranges and choose the values of $\delta \rho/ \rho$=80
($D_0$=0.41 Mpc) and $cz_0$=200 km/s (corresponding to 234 km/s at the
velocity of 1000 km/s) for our final percolation catalog with customary
scalings of the two search parameters.  According to eq.  (3), $D_L$ is
0.48, 0.61, 0.89, 1.05, and 1.27 Mpc at 1000, 2000, 4000, 5000, and 6000
km/s, respectively, whereas $cz_L$ is 234, 298, 434, 519, and 620 km/s at
the respective distances.  The resulting catalog turns out to be in 
good agreement with that obtained with the H algorithm (see \S 5).

The choice of a less restrictive velocity link parameter would lead to
group catalogs more dissimilar to that of hierarchical groups, i.e. with
an even smaller fraction of ungrouped galaxies and binary pairs and an
even larger number of groups.  For instance, choosing $cz_0=$300 km/s
and the same value of $D_0$, we obtain a 7\% smaller number of ungrouped
galaxies, a 4\% smaller number of binary pairs, and a 3\% greater number
of systems with at least three members. On the other side, choosing
$cz_0=$100 km/s and the same value of $D_0$, we obtain twice the number
of ungrouped galaxies, together with only about 1/6 of the groups with
at least three members. If we let $\delta \rho/\rho$ decrease to 60 
(increase to 100), with $cz_0$=200 km/s, we obtain 8\% less (6\% more) 
ungrouped galaxies; the numbers of galaxy pairs and systems with at 
least three members vary by a smaller percentage in the same and 
opposite sense, respectively. 
 
Several simulations (Nolthenius \& White 1987, Moore, Frenk \& White 1993,
Nolthenius, Klypin \& Primack 1994, 1997) suggest that the
above-mentioned scaling of the velocity link parameter $cz_L$ increases
too rapidly at large redshifts (see also Nolthenius 1993) and favour a
mild increase of $cz_L$ with $z$ (together with a similar scaling of
$D_L$) from about 200--400 km/s at 500 km/s to about 400--700 km/s at 6000
km/s, with details (especially the zero-point of the scaling relation)
depending on the adopted cosmological model. A mild scaling of $cz_L$ 
with $z$ has the advantage of minimizing the number of interlopers at the 
price of failing to pick up all members of clusters characterized by 
high velocity dispersion (see, e.g., Nolthenius 1993; Frederic 1995 a, 
b).  

In the absence of compelling reasons for making a precise choice of the detailed
scaling of $cz_L$, we have run the P algorithm also keeping $cz_L$ constant with
$z$, i.e. $cz_L=cz_0$ (and $D_L$ scaled as above). This is an extreme choice which,
though conceptually very questionable, is used here in practice as an approximation
to a slow variation of $cz_L$ with $z$, given the limited range of $z$ encompassed
by NOG. Also Garcia (1993) used the same approximation (i.e. $cz_L$ constant) in her
application of the P algorithm to a sample of nearby galaxies limited to the depth
of 5500 km/s. 
 
We have run the P algorithm for some pairs of values of the two linking
parameters lying in the above-mentioned ranges and we choose the values
of $\delta \rho/\rho$=80 ($D_0$=0.41 Mpc) and $cz_L$=350 km/s for our
final P group catalog with $cz_L$ kept constant. 

If we let $\delta \rho/\rho$ decrease to 60 (increase to 100), with
$cz_L$=350 km/s, the fraction of ungrouped galaxies decreases by 8\%
(increases by 6\%) and the number of galaxy pairs accordingly varies by a
smaller percentage. On the other hand, if we let $cz_L$ vary from
$cz_L$=250 km/s to 600 km/s, with $\delta \rho/\rho$=80, the number of
ungrouped galaxies decreases from values 10\% greater to values 10\%
smaller than that relative to $cz_L$=350 km/s; the number of pairs
accordingly varies by a smaller percentage. The number of groups with at
least three members does not change appreciably in all these cases. 

The two variants of the P algorithm (with $cz_L$ kept constant and with
$cz_L$ scaled with $z$) considered in this paper are meant to represent
two extreme cases for the scaling behaviour of $cz_L$. As discussed in
\S 5, it is encouraging that the two respective catalogs of groups,
hereafter denoted as P1 and P2 respectively, appear to be in very close
agreement between each other; they turn out to be also in good agreement
with the catalog of H groups, with P1 in sligthly better agreement than
P2. Clearly, for our sample which covers a limited range of distances,
differences in the adopted scaling of the velocity link parameter of the
P algorithm are unimportant. 

In each of its variants, the P algorithm groups together many nearby galaxies
(among them many members of the Virgo and Ursa Major clusters and of well-known
very nearby groups) into a very large unrealistic system, even if we let the
values of the parameters $cz_0$ and $\delta\rho/\rho$ vary within reasonable
intervals. Garcia (1993) encountered a similar problem in running the P algorithm
for her sample of comparatively nearby galaxies. This problem stands out when the
algorithm is applied to a dense sample of nearby galaxies. The problem is mainly
related to the fact that the galaxies which at a given step are merged into a
group are picked up only in reference to their closest neighbour in the group and
not to the whole set of galaxy members gathered at the previous steps (as is done
in the case of the H algorithm). This can lead to sort out possible non-physical
systems, like a long filament of galaxies with a small separation between
physically unrelated neighbouring objects. 

We have solved this problem by taking directly a few very nearby groups
and the systems of the Virgo region as given in the literature (as
explained at the end of \S 5.1) and by adopting the same results obtained
with the H method in the nearby region ($cz<500$ km/s).  Therefore, by
definition the catalogs of groups selected with the P method are equal to the
catalog of H groups in the Virgo region and nearby region ($cz<$500
km/s). 

\section{The catalogs of groups}

Although we have identified groups in  redshift space, we expect  
the group selection to be hardly affected by peculiar motions, since 
all galaxies located in a small volume tend to move together in redshift 
space.  

Our final catalog of H groups comprises 1062 systems, i.e.  587 binaries and 475
groups with at least three members. These groups contain 3119 galaxies. Of
these groups 413 comprise $n<10$ members for a total of 1723 galaxies, 39
groups comprise $10\leq n < 20$ members for a total of 494) galaxies, and
23 groups (among which the major Virgo substructures and the well-known
clusters Ursa Maior, Fornax, Eridanus, Centaurus, Hydra) have at $n\geq$20
members for a total of 902 galaxies.  The remaining 2783 galaxies are left
ungrouped (field galaxies). 

Our final catalog of P1 (P2) groups comprises 1079 (1093) systems, i.e. 572 (581)
binaries and 507 (512) groups with at least three members. These groups contain
3239 (3295) galaxies; of them 444 (448) groups comprise less than 10 members
for a total of 1842 (1889) galaxies, 44 (45) groups comprise $10\leq n < 20$
members for a total of 580 (587) galaxies, and 19 (20) groups have at least 20
members for a total of 817 (819) galaxies. There are 2693 (2619) galaxies which
are left ungrouped (field galaxies). 

Table 1 shows the numbers of H, P1, P2 groups for different group richness
(number $n$ of galaxy members). By applying the nonparametric
Kolmogorov-Smirnov and sign statistical tests (e.g., Hoel 1971), we find no
significant differences between the distributions. Thus, the three catalogs of
groups are, on average, similar as far as the distribution of galaxy members in
groups is concerned.

Furthermore, we quantify the similarity between the catalogs of groups by counting
the number of members of a H group which belong to a common P1 group. We first
determine which members of each P1 group belong to the same H group. We calculate
a largest group fraction (LGF) for each P1 group by dividing the number of members
in the largest such subgroup by the total number of members in the P1 group (see
Frederic 1995a for a similar definition of LGF).  Fig. 5 shows, as a function of
group richness (number of members), the fraction of P1 groups of a given richness
with LGFs of unity and in each quartile below. For example, there are 22 P1 groups
with seven members. Of these, 48\% have LGF of 100\%, 57\% have LGF of 75\%, 91\%
have LGF of at least 50\%, and all of the n=7 groups have LGFs greater than 25\%. 
The H groups give a similar histogram, with somewhat greater values along the
ordinate axis (see Fig. 6). The large fractions of groups having high LGF-values
confirm the similarity between the two catalogs of groups. If we repeat these
calculations replacing P1 groups with P2 groups, we find slighly lower values
along the ordinate axis in the plot corresponding to Fig. 5 and an almost equal
histogram in the plot corresponding to Fig. 6. Thus, P1 groups are in slightly
better agreement with H groups than P2 groups.  If we compare P1 and P2 groups in
the same way, we find a very good agreement, as expected (the values of LGF are
almost always greater than 80\% and are frequently greater than 90\%). 
 
Furthermore, we have calculated the LGF-values separately for the nearby and distant 
NOG galaxies dividing the sample at 3500 km/s. In this way, we have verified that 
the agreement between the P1 and P2 groups gets slightly worse as we go to larger distances, as 
expected. On the other hand, there is no appreciable effect of this kind in the comparison 
between H and P1 (or P2) groups. 

The ratio of the number of groups with at least three members to the
number of non-member (binary and field) galaxies is 0.12, 0.13, 0.14 for
the H, P1, P2 groups, respectively.  These values lie in the range of
published values coming from other group catalogs, e.g., 0.09 for the
SSRS1 groups (Maia et al. 1989), 0.10 for the LCRS groups (Tucker et al.
1997) and the PPS groups (Trasarti--Battistoni 1998), 0.11 for the PGC
groups (Gourgoulhon et al. 1992, Fouqu\'e et al. 1992) groups, 0.12 for
the SSRS2 groups (Ramella et al. 1999b), 0.13 for the CfA2 north
(Ramella et al. 1997) and ESP groups (Ramella et al. 1999a), 0.14 for
the revised CfA1 groups (Nolthenius 1993), 0.17 for the NBG groups
(Tully 1987), 0.15 and 0.19 for the LEDA groups derived by Garcia (1993)
using the P and H methods, respectively. 

The ratio of members of groups with at least three members to the total
number of galaxies is 0.44, 0.46, 0.47 for the H, P1, P2 groups,
respectively, whereas published values are 0.35 for the SSRS1 (Maia et
al. 1989), LCRS (Tucker et al. 1997) and PPS groups
(Trasarti--Battistoni 1998), 0.40 for the SSRS2 (Ramella et al. 1999b)
groups, 0.41 for the ESP groups (Ramella et al. 1999a) groups, 0.42 for
the PGC groups (Gourgoulhon et al. 1992;  Fouqu\'e et al. 1992), 0.45
for the CfA2 north groups (Ramella et al. 1997), 0.48 for the revised
CfA1 groups (Nolthenius 1993), 0.51 for the NBG groups (Tully 1987),
0.63 and 0.47 for the LEDA groups (Garcia 1993), respectively derived by
means of the P and H methods. 

In general, our catalogs of groups are broadly consistent with the previous
catalogs of groups selected in the same regions and our values for the two
above-mentioned ratios appear to be consistent with typical values reported in
the literature. 

As regards the H groups, our values are close to those of the PGC groups
and are a little lower than those of the NBG groups (because we adopt a
greater limiting luminosity density parameter to cut the hierarchy (see \S
4.1)). Compared to the LEDA groups identified with the H method, we find
less groups, which is partially due to the fact that in many cases Garcia
(1993) neglected the reconstruction of high-velocity systems, which the
algorithm tends to break in several systems with different average
velocities (see \S 4.1).  Furthermore, compared to the LEDA groups
identified with the P method, we basically find smaller groups with less
members, because, on average, we adopt lower values of $cz_L$ (see \S 4.2).  In
general, there is much less similarity between the two catalogs of LEDA
groups than between our two catalogs. 

A comparison of the distribution of the centers of the two samples of
groups with that of galaxies show qualitatively that groups trace the
large-scale structure of the nearby universe. 

The final catalogs of the members of H, P1, and P2 groups are presented in
Tables 2, 3, and 4, respectively. In these Tables we give the number of group,
the PGC and alternative names of the galaxy member, the 1950 right ascension
and declination (in hours, minutes, seconds and in degrees, arcmin, arcsec,
respectively), the velocity $cz$ (in the Local Group frame), and the corrected
total blue magnitude. 

The final catalogs of H, P1, and P2 groups (along with some group properties)
are presented in Tables 5, 6, and 7, respectively. These tables give the NOG
group number, the name of the brightest galaxy of the group, the number of
galaxy members, the median values of the 1950 right ascension and declination
of the group members, the median value of recession velocity $cz$ (in the Local
Group frame), the common name of the system (when available), the
cross-identifications between NOG groups, the cross-identification
between NOG groups and previous catalogs of groups. Of them we choose the
all-sky catalogs of nearby groups published by Tully (1987) and by Garcia
(1993) for a detailed comparison. Specifically, we consider Garcia' s (1993)
final catalog of groups defined by her as the one that includes only systems
common to the two original catalogs that she constructed by means of the H and
P methods.  Cross-identifications are tabulated only when there at least three
galaxies in common between our groups (with at least three members) and groups
of previous catalogs and two galaxies in common between pairs. 

In Table 5 we denote by an asterisk the 17 systems which are split by the H
algorithm along the line of sight and then are reconstructed by us with
the aid of the results of the P1 method.  Moreover, in Table 5 we denote by
a flag $+$ the 11 systems which are constructed with the aid of membership
assignments provided directly in the literature for the Virgo region
(seven systems and 311 galaxies) and for four very nearby groups
(comprising 55 galaxies) (see \S 4.1). As explained at the end of \S 4.2,
the P1 and P2 systems are by definition taken to be
equal to those identified with the H method in the Virgo region and in the
very nearby region ($cz<$500 km/s). The latter region involves 13 systems
(of which 3 pairs) and 161 (118 grouped and 43 ungrouped) galaxies. These systems
are denoted by a flag $+$ in Tables 6 and 7.

Tables 2, 3, 4, 5, 6, and 7 are available in electronic form only. 

\section{Conclusions}

In this paper we describe the NOG sample, a distance-limited ($cz<$6000 km/s)
and magnitude-limited (B$\leq$14 mag) sample of 7076 optically-selected
galaxies which covers 2/3 of the sky ($|b|>20^{\circ}$) and has a good
completeness in redshift (98\%). 
 
We select the NOG on the basis of homogenized corrected blue magnitudes in
order to minimize systematic effects in galaxy sampling, due to the use of
different magnitude systems in different areas of the sky and to Galactic
and internal extinction. In this sense the NOG, which is meant to be the
first step towards the construction of a statistically well-controlled
optical galaxy sample with homogenized photometric data covering most of the
celestial sphere, is in principle designed to offer a largely unbiased
view of the galaxy distribution. 

We identify galaxy systems in the NOG by means of both the hierarchical and the
percolation {\it friends of friends} methods.  After an extensive search in the
space of relevant parameters with the guide of available numerical simulations, we
choose optimal sets of parameters which allow us to obtain reliable and
homogeneous catalogs of loose groups. Remarkably, these catalogs turn out to be
substantially consistent as far as the distribution of members in groups is
concerned.  Containing about 500 systems (with at least three members), they are
among the largest catalogs of groups presently available. Although they are drawn
from a galaxy sample limited to bright magnitudes, they are useful for studies of
the statistical properties of loose groups, since their physical properties were
found to be stable, on average, against the inclusion of fainter galaxy members
(Ramella et al. 1995a,b; Ramella, Focardi \& Geller 1996). In particular, being
extracted from the same galaxy sample, the catalogs allow one to investigate on
variations in group properties (e.g., velocity dispersion, virial mass and radius)
strictly related to differences in the algorithm adopted. These differences
indicate to what extent our knowledge of the location and properties of groups in
the nearby universe is inaccurate.  Previous comparisons between catalogs of
groups identified with the H and P algorithms (Pisani et al. 1992) were based on
catalogs extracted from different galaxy samples. 

Most of the NOG galaxies ($\sim$60\%) are found to be members of galaxy
pairs ($\sim$580 pairs comprising $\sim$15\% of the galaxies) or groups
with at least three members ($\sim$500 groups comprising $\sim$45\% of the
galaxies). About $\sim$40\% of the galaxies are left ungrouped (field
galaxies). 

Though being limited to a depth of 6000 \ks, the NOG covers interesting
regions of prominent overdensities (in mass and galaxies) of the nearby
universe, such as the "Great Attractor" region and the Perseus-Pisces
supercluster.  Compared to previous all-sky optical and IRAS galaxy
samples, the NOG provides a denser sampling of the galaxy density field in
the nearby universe. Besides, as expected, the NOG delineates overdensity 
regions with a greater density contrast than IRAS galaxy samples do. 

Given its high-density sampling and large sky coverage, the NOG sample is
well suited for mapping the cosmography of the nearby universe beyond the
Local Supercluster and for allowing a comparison of the density field as
traced by optical galaxies with that described by IRAS galaxies
(addressing questions concerning the amount of relative biasing in the
galaxy distribution and its possible dependence on scale). 

By virtue of the identification of NOG groups, the NOG is also well suited
for deriving galaxy density parameters on small scales to be used in
observational investigations of environmental effects on galaxy
properties. Environmental studies in which the local galaxy
density is decoupled from membership in galaxy systems go beyond the
conventional comparison between the properties of cluster and field
galaxies and thus can better constrain physical processes responsible for
the formation and evolution of galaxies. Much of the observed evolution of
the properties and populations of galaxies (e.g., Ellis 1997) which has
occurred during recent epochs ($z<1$) can be ascribed to interaction of
galaxies and their local environment. 

In a subsequent paper (see Marinoni et al. 1999b for preliminary results) the
NOG groups will be used to remove non-linearities in the peculiar velocity
field (e.g., the velocity dispersion of group members) on small scales. To
correct the redshift--distances of field galaxies and groups on large scales,
we shall apply models of the peculiar velocity field, following the approach
described in Paper I.  We shall use the locations of individual
galaxies and groups calculated in real--distance space (i.e. for distances
predicted by different velocity field models) to calculate the selection
function of the NOG sample (see Paper II) and to reconstruct the galaxy density field. 
Local galaxy density parameters to be used in studies of environmental effects on
nearby galaxies will be provided.

\acknowledgements

We are indebted to B. Santiago (together with the ORS team) and to 
W. Saunders (together with the the PSCz team) who provided us with data 
in advance of publication.

We wish to thank S. Borgani, D. Fadda, R. Giovanelli, M. Girardi, M. Hudson, F.
Mardirossian, M.  Mezzetti, P. Monaco, M.  Ramella for interesting conversations. C. 
M. and L. C. are grateful to SISSA for its kind hospitality. 

This research has made use of the Lyon-Meudon Extragalactic Database 
(LEDA) supplied by the LEDA team at the CRAL-Observatoire de Lyon 
(France) and of the NASA/IPAC Extragalactic Database (NED) which is 
operated by the Jet Propulsion Laboratory, California Institute of 
Technology, under contract with the National Aeronautics and Space 
Administration.

This work has been partially supported by the Italian Ministry of 
University, Scientific and Technological Research (MURST) and by the 
Italian Space Agency (ASI).

\newpage

\figcaption[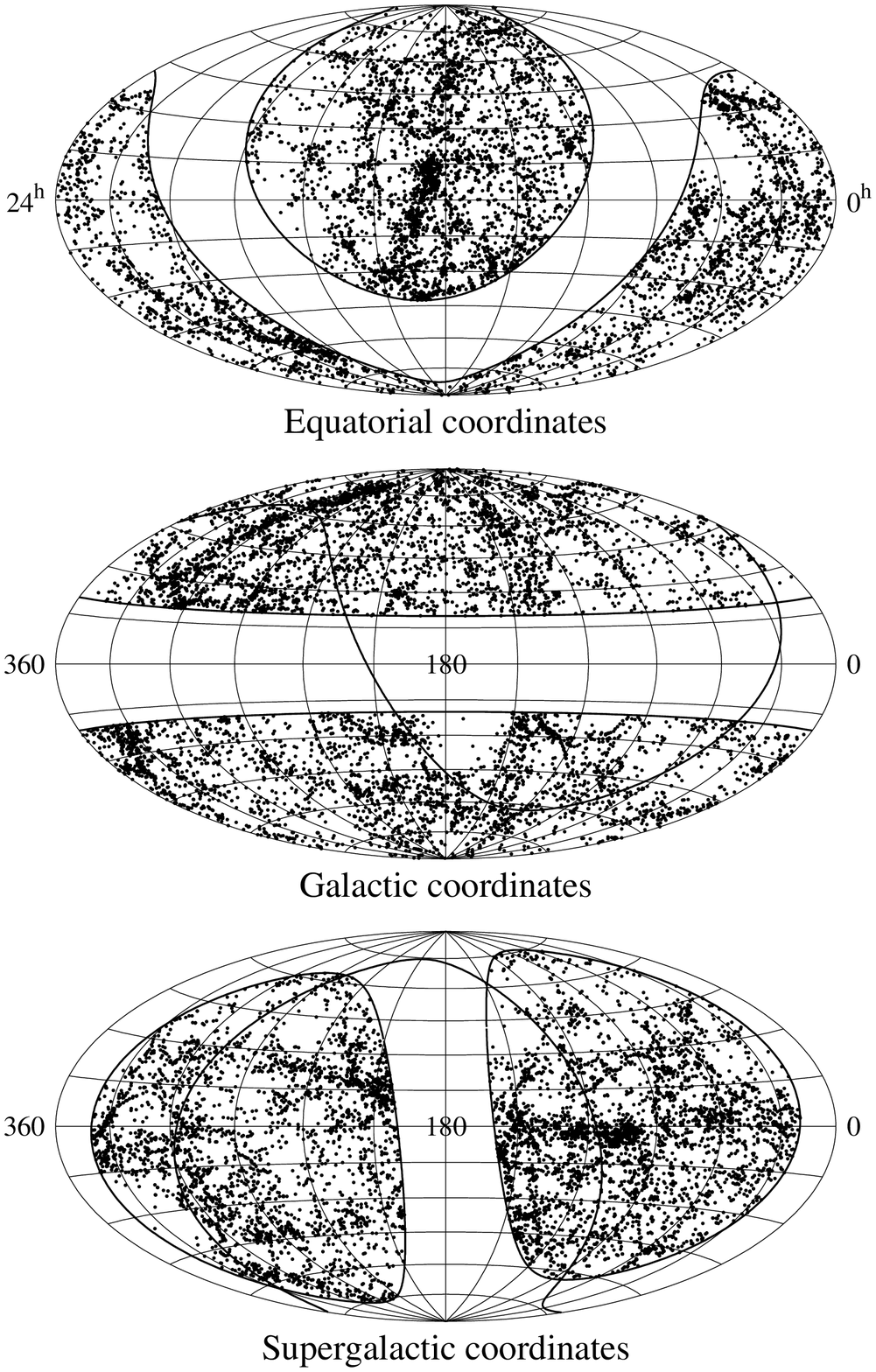]{The NOG sample is shown in equal-area
Aitoff projections on the sky using equatorial, Galactic, and
supergalactic coordinates. The region devoid of galaxies is the zone of
avoidance ($|b|<20^{\circ}$). The heavy line is drawn at the celestial
equator, $\delta=0^{\circ}$.}

\figcaption[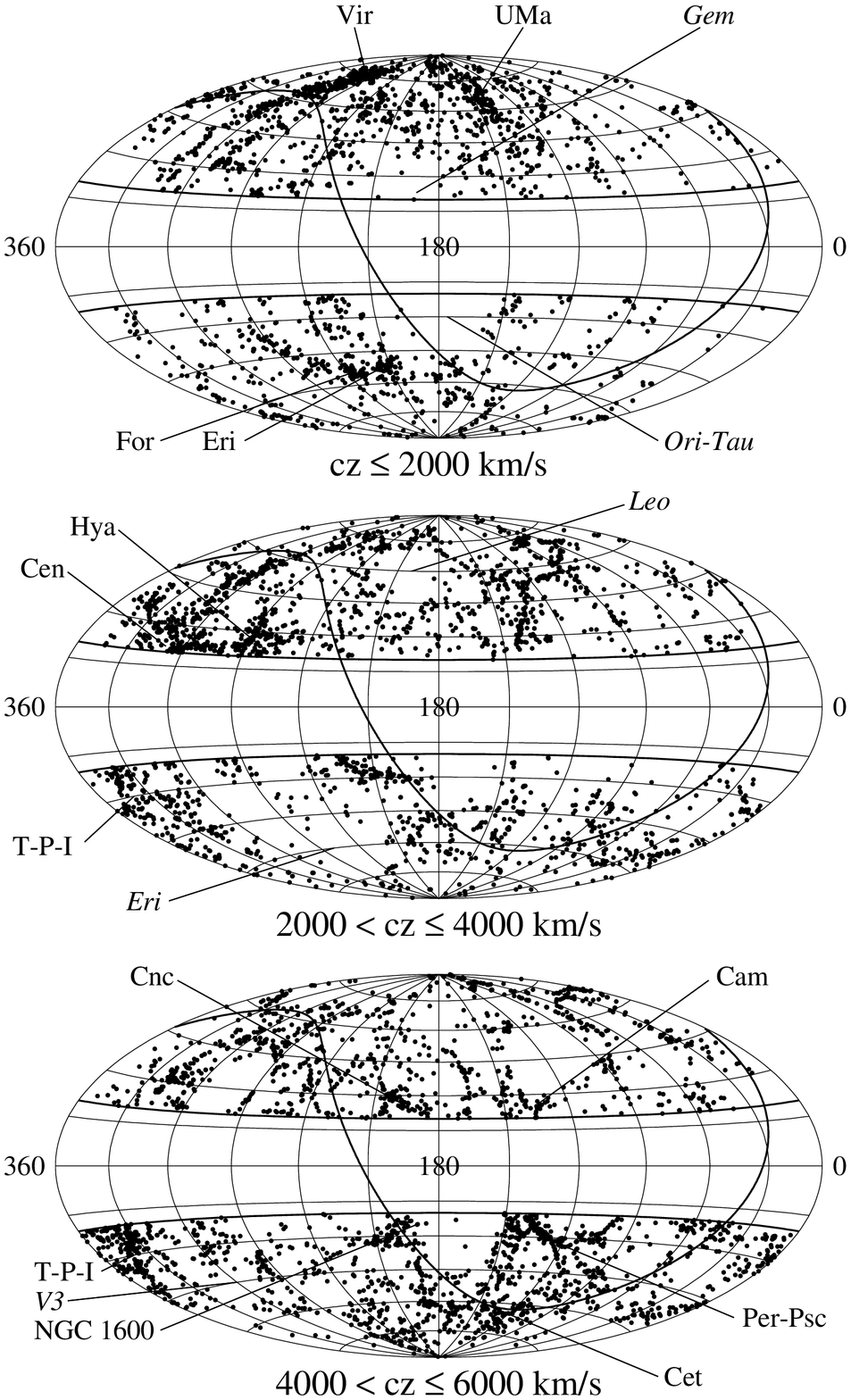]{The NOG sample is shown in equal-area Aitoff
projections on the sky using Galactic coordinates, for three different
redshift slices. The region devoid of galaxies is the zone of avoidance
($|b|<20^{\circ}$).  The S-shaped line is drawn at the celestial equator,
$\delta=0^{\circ}$. Several major structures and voids mentioned in the text 
are marked. Voids are marked in italics.}

\figcaption[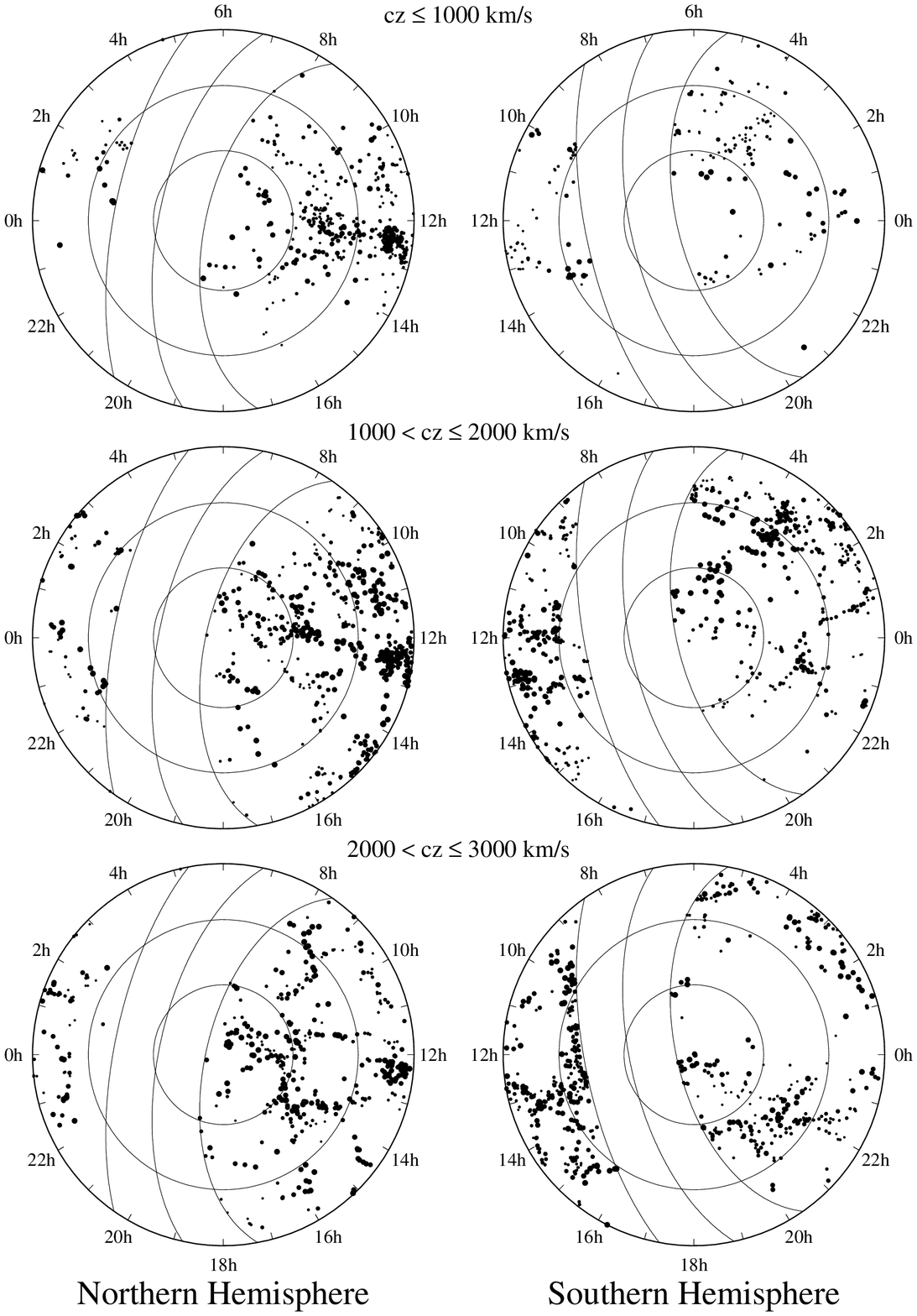]{The NOG sample is shown in equal-area polar
hemispheric projections in equatorial coordinates, for the three redshift
slices indicated. The circle on the left (right) side corresponds to the
north (south) celestial hemisphere. The poles are at the center of these
circles with celestial latitude decreasing radially outward; circular
lines are drawn at declinations $|\delta|=30^{\circ}$ and
$|\delta|=60^{\circ}$. Right ascension runs azimuthally as indicated. The
region devoid of galaxies is the zone of avoidance ($|b|<20^{\circ}$).}

\figcaption[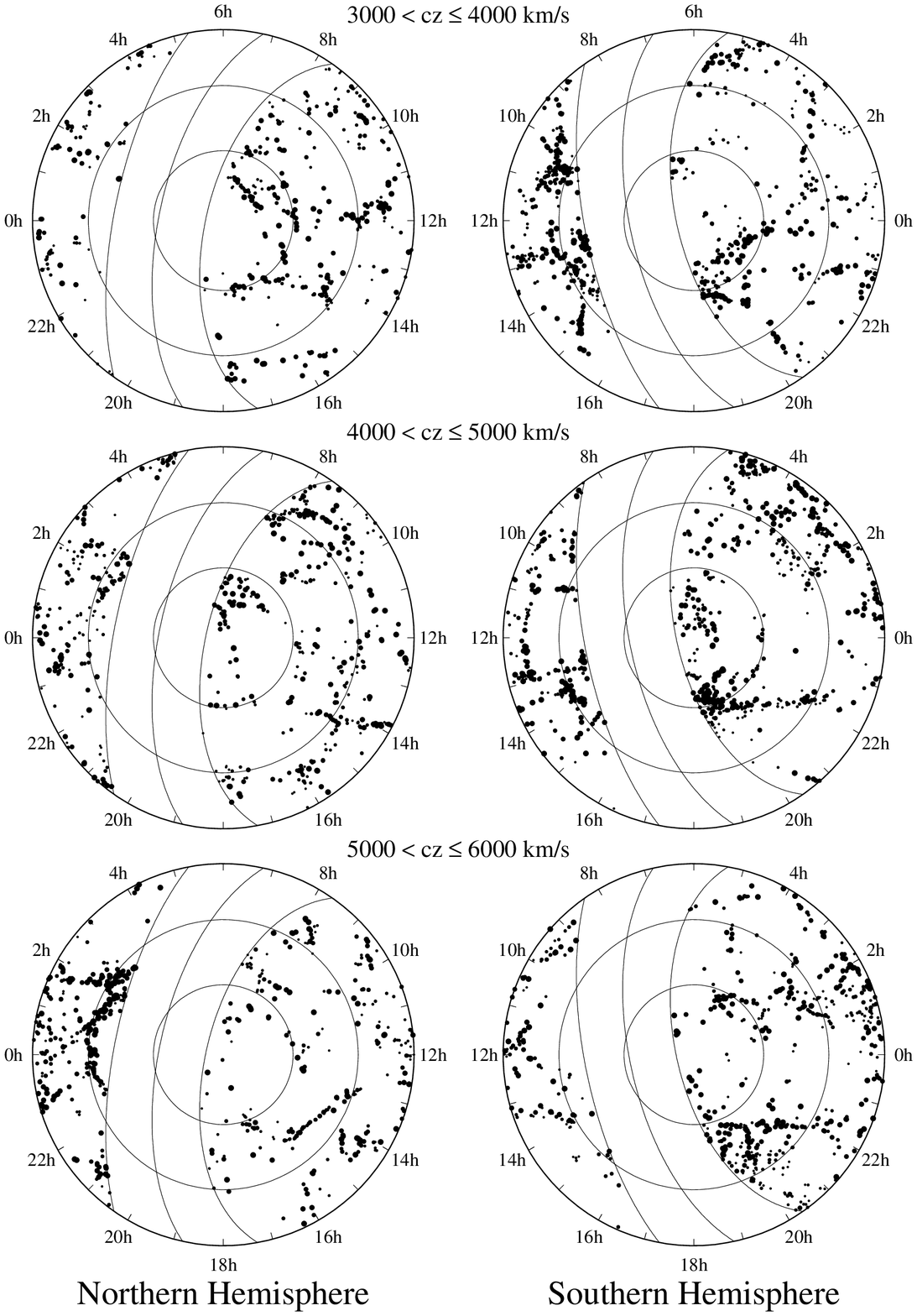]{The NOG sample is shown in equal-area polar
hemispheric projections in equatorial coordinates, for the three redshift
slices indicated. Lines as in Fig. 3.}

\figcaption[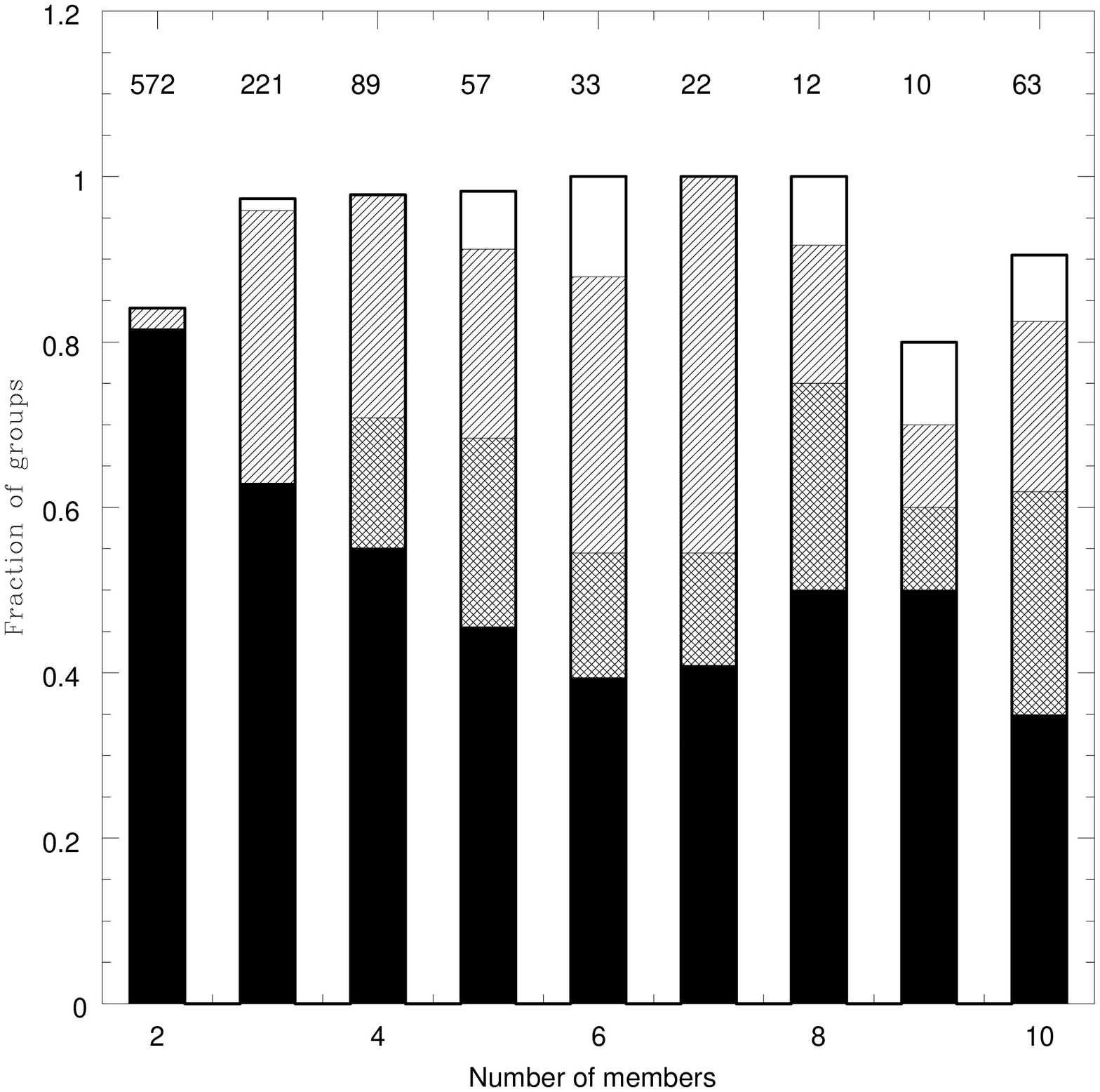]{Histogram of the largest group fraction (LGF) as a function
of the number of galaxy members in P1 groups.  Black-hatched regions give the
fraction of P1 groups with LGFs of unity, cross-hatched regions correspond to
groups with LGF between 75\% and 100\%, single narrow-hatched regions correspond
to LGFs between 50\% and 75\%, and no hatching represents groups with LGFs between
25\% and 50\%. The number at the top of each bar is the total number of P1 groups
with the given number of members n. The n=10 bar includes all groups with 10 or
more members.}

\figcaption[]{Histogram of the largest group fraction (LGF) 
as a function of the number of galaxy members in H groups. Hatching as 
in Fig. 5.} 

\clearpage

\begin{deluxetable}{lrrr}
\tablewidth{0pc}
\tablecaption{The distribution of galaxy members in NOG groups}
\tablehead{
\colhead{n}  &
\colhead{H}  &
\colhead{P1} &
\colhead{P2}\\
}
\startdata 
n=2           & 587 & 572 & 581\\
n=3           & 194 & 221 & 212\\
n=4           &  91 &  89 &  93\\
n=5           &  55 &  57 &  65\\
n=6           &  32 &  33 &  31\\
n=7           &  23 &  22 &  20\\
n=8           &  13 &  12 &  13\\
n=9           &   5 &  10 &  14\\
10$\le$n$<$20 &  39 &  44 &  45\\  
n$\ge20$      &  23 &  19 &  20\\

\enddata
\end{deluxetable}

\clearpage

\newpage

\begin{figure}
\centerline{
\psfig{figure=fig1.eps,height=20cm,angle=0}
}
\end{figure}
\clearpage

\begin{figure}
\centerline{
\psfig{figure=fig2.eps,height=20cm,angle=0}
}
\end{figure}
\clearpage

\begin{figure}
\centerline{
\psfig{figure=fig3.eps,height=20cm,angle=0}
}
\end{figure}
\clearpage

\begin{figure}
\centerline{
\psfig{figure=fig4.eps,height=20cm,angle=0}
}
\end{figure}
\clearpage

\begin{figure}
\centerline{
\psfig{figure=fig5.eps,height=20cm,angle=0}
}
\end{figure}
\clearpage

\begin{figure}
\centerline{
\psfig{figure=fig6.eps,height=20cm,angle=0}
}
\end{figure}

\end{document}